\documentclass[journal]{IEEEtran}

\usepackage{graphicx}
\usepackage{amsmath}
\usepackage{amssymb}
\usepackage{amsthm}
\usepackage{mathtools}
\usepackage{bm}
\usepackage{subcaption}
\usepackage{xfrac}
\usepackage{pgfplots}
\pgfplotsset{compat=1.14} 
\usepackage{xargs}
\usepackage{cite}
\usepackage{url}

\allowdisplaybreaks 

\newtheorem{assump}{Assumption}
\newtheorem{remark}{Remark}
\newtheorem{lemma}{Lemma}
\newtheorem{theorem}{Theorem}
\newtheorem{definition}{Definition}
\newtheorem{corollary}{Corollary}

\begin{document}

\title{Interior Point Differential Dynamic Programming}

\author{Andrei~Pavlov,~\IEEEmembership{Student Member,~IEEE,}
        Iman~Shames,~\IEEEmembership{Member,~IEEE,}
        and~Chris~Manzie,~\IEEEmembership{Senior Member,~IEEE.}
\thanks{The authors are with the Department
of Electrical and Electronic Engineering, The University of Melbourne, Parkville,
Victoria, 3010 Australia, e-mail: {\tt \{apavlov@student.,ishames@,manziec@\}unimelb.edu.au}}
}
\maketitle

\begin{abstract}
This paper introduces a novel Differential Dynamic Programming (DDP) algorithm for solving discrete-time finite-horizon optimal control problems with inequality constraints. Two variants, namely Feasible- and Infeasible-IPDDP algorithms, are developed using primal-dual interior-point methodology, and their local quadratic convergence properties are characterised. We show that the stationary points of the algorithms are the perturbed KKT points, and thus can be moved arbitrarily close to a locally optimal solution. Being free from the burden of the active-set methods, it can handle nonlinear state and input inequality constraints without a discernible increase in its computational complexity relative to the unconstrained case. The performance of the proposed algorithms is demonstrated using numerical experiments on three different problems: control-limited inverted pendulum, car-parking, and unicycle motion control and obstacle avoidance.
\end{abstract}


%
\IEEEpeerreviewmaketitle

\maketitle


\section{Introduction}

Among optimisation algorithms that can effectively utilise the structure of the optimal control problems a distinct place is held by Differential Dynamic Programming (DDP) algorithm introduced by Mayne\cite{mayne1966second}. Its advantages include linear complexity in the length of prediction horizon and locally optimal feedback policies as an extra output. It has provable local quadratic convergence \cite{de1988differential}, while global convergence with inexact line-search can also be established\cite{liao1991convergence}. DDP can address minimax-type problems \cite{morimoto2003minimax} or be used for Model Predictive Control applications\cite{tassa2008receding}. However, the DDP approach requires second-order derivatives of system dynamics, which makes it prohibitively expensive for solving large problems. As a possible remedy, one can omit the second-order information as in iterative-Linear-Quadratic Regulator (iLQR)\cite{li2004iterative}, or estimate it using Quasi-Newton DDP\cite{sen1987quasi}, Sampled DDP\cite{rajamaki2016sampled} or Unscented DDP\cite{manchester2016derivative}.

One major shortcoming of DDP is a lack of an elegant generalisation for the inequality-constrained problems. Methods reported in the literature usually fall into one of two distinct categories: penalty, barrier and Augmented Lagrangian (AL) methods; or active sets methods. The first family utilises penalty and barrier functions (or augments the Lagrangian function) to convert constrained problems into unconstrained, e.g., see \cite{pellegrini2017applications}\cite{plancher2017constrained}. Algorithms of the first family potentially suffer from major drawbacks, such as ill-conditioning, need for hand-tuning, slow convergence and/or appearance of saddle-points. The second family is based on active-set methodology, e.g., see \cite{murray1979constrained,yakowitz1986stagewise,lin1991differential,xie2017differential}. In contrast to the former family, these methods require the constraints to be explicitly dependent on the control variables, and include an extra routine aimed at identification of active/inactive constraints. This approach is known to have combinatorial complexity in the number of constraints in the worst case. To circumvent potential computational difficulties Tassa et. al.\cite{tassa2014control} propose Control-Limited Differential Dynamic Programming (CLDDP) algorithm where only box constraints on input are considered. A hybrid algorithm which uses ideas from both families was reported in \cite{lantoine2012hybrid}. To the best of our knowledge there are no rigorous proofs of convergence for constrained DDP algorithms for problems with nonlinear state and input constraints.  

While general purpose primal-dual interior-point methods are particularly successful in practice (see \cite{gondzio2012interior} for details), there were no extension of DDP algorithm accommodating primal-dual interior-point techniques reported so far. Our aim is to fill the gap and bring the benefits of this approach. As demonstrated in the following, the primal-dual Interior-Point DDP algorithm seems to be one of the most natural extensions to DDP. It requires neither modifying the objective function nor identifying active/inactive constraints by a separate procedure. Importantly, it has provable local quadratic convergence for problems with nonlinear state and control constraints, which is a new result in the DDP framework.

The paper is organised as follows. In Section \ref{sec:prelim}, we introduce the finite-time optimal control problem, and state a system of (perturbed) first-order conditions for local optimality. In Section \ref{sec:alg}, the main contribution of the paper, i.e., two extensions for the Differential Dynamic Programming (DDP) algorithm, namely Feasible- and Infeasible-Interior-Point DDP algorithms are described and their local quadratic convergence properties are established. In Section \ref{sec:examples} we consider numerical examples to demonstrate advantages in convergence relative to existing approaches (CLDDP and log-barrier DDP).

\section{Preliminaries}\label{sec:prelim}
    Consider a discrete-time system of the following form
    \begin{equation*} \label{sys}
        x_{t+1}= f(x_t,u_t),
    \end{equation*}
subject to constraints $c(x_t,u_t)\leq 0$,
where $x_t\in \mathbb{R}^n$ and $u_t\in \mathbb{R}^m$ are state and control vectors at time $t$, functions $f:\mathbb{R}^n\times \mathbb{R}^m \rightarrow \mathbb{R}^n$ and $c:\mathbb{R}^n\times \mathbb{R}^m \rightarrow \mathbb{R}^l$ are twice continuously differentiable.

Consider a finite-time constrained optimal control problem for the above system at a known initial state $\bar x_0$:
\begin{equation} \label{CFTOCP}
\begin{aligned}
J^{\star}_N(\bar x_0)=&\min_{\bm x, \bm u} & &\sum_{t=0}^{N-1} q(x_t, u_t) + p(x_{N}) \\
&\text{\ s.t.} & &x_0=\bar x_0,\text{ and for }t\in\{0,\ldots,N-1\}:\\
& & &x_{t+1}=f(x_t, u_t),\ c(x_t, u_t)\leq 0, 
\end{aligned}
\end{equation}
where $q:\mathbb{R}^n\times \mathbb{R}^m \rightarrow \mathbb{R}$ and $p:\mathbb{R}^n \rightarrow \mathbb{R}$ are twice continuously differentiable stage and terminal costs respectively,  $N\geq 1$ is the prediction horizon length, $\bm x$ and $\bm u$ are vectors of the corresponding decision variables, i.e., $\bm x=(x_0,\ldots,x_N)$ and $\bm u=(u_0,\ldots,u_{N-1})$.


The Lagrangian function for this optimisation problem is
\begin{multline*}\label{Lagrange}
    L(\bm x, \bm u, \bm \lambda,  \bm s) = \sum_{t=0}^{N-1} q(x_t, u_t)+ s_t^T  c(x_t, u_t)+p(x_N)\\ +\lambda_0^T( \bar x_0-x_0)
    + \sum_{t=1}^{N}\lambda_{t}^T  (f(x_{t-1},u_{t-1})-x_{t}),
\end{multline*}
where $s_t\in \mathbb{R}^l$ and $\lambda_t\in \mathbb{R}^n$ are the dual variables. 

Let $\bm c(\bm x, \bm u)=\big(c(x_0,u_0),\ldots,c(x_{N-1},u_{N-1})\big)$, $\bm \lambda=(\lambda_0,\ldots,\lambda_N)$, $\bm s=(s_0,\ldots,s_{N-1})$ and $\bm S=\operatorname{diag}[\bm s]$. The perturbed KKT system is defined as
\begin{equation}\label{KKT}
\begin{aligned}
    \nabla_{\bm x} L(\bm x,\bm u,\bm \lambda, \bm s) &= 0,\\
    \nabla_{\bm u} L(\bm x,\bm u,\bm \lambda, \bm s) &= 0,\\
    \nabla_{\bm \lambda} L(\bm x,\bm u,\bm \lambda, \bm s) &= 0,\\
    \bm S \bm c(\bm x, \bm u) +\bm \mu&=0,\\
    \bm c(\bm x, \bm u)\leq0,\ \bm s&\geq 0,
\end{aligned}
\end{equation}
where $\nabla$ is the gradient operator, inequalities are understood to be element-wise, and $\bm \mu$ is a vector of perturbations $\mu>0$ of an appropriate dimension. The KKT system, i.e., system \eqref{KKT} with $\mu =0$, defines the first-order necessary conditions for a local constrained minimiser.
\begin{assump} \label{as:0}
The KKT system has a solution $(\bm x^{\star}, \bm u^{\star},\bm \lambda^{\star}, \bm s^{\star})$, which satisfies the following conditions:
\begin{enumerate}
    \item Strict complementary holds at the solution, i.e., $c(x_t^{\star}, u_t^{\star})<s^{\star}_t$ for $t\in\{0,\ldots,N-1\}$. 
    \item The standard second-order constrained optimality conditions\footnote{Here we refer to the conditions of the type outlined in \cite[Theorem 12.6]{nocedal2006numerical}}. hold at $(\bm x^{\star}, \bm u^{\star},\bm \lambda^{\star}, \bm s^{\star})$.
\end{enumerate}
\end{assump}
Under Assumption~\ref{as:0} and some additional regularity requirements, the optimal solution of \eqref{CFTOCP} can be obtained as the limit point of the solutions to the perturbed KKT system for decaying perturbation, e.g., see \cite{el1996formulation}.
\begin{remark}[Iteration Index Convention]
For the clarity of notation, we drop decision variables' iteration indices. Instead we use the superscript ${}^+$ to denote the value of a variable at the next iteration. A variable's value at the current iteration is denoted by the variable name without this superscript.
\end{remark}


\section{Primal-Dual Interior-Point Differential Dynamic Programming}\label{sec:alg}
In this section we introduce the primal-dual Interior-Point Differential Dynamic Programming (IPDDP) algorithms for solving inequality constrained optimal control problems\footnote{As no time invariance properties are required, the proposed algorithms can be applied to the case where $f$, $c$ and $q$ functions are time-varying.}. Note that equality constraints can be addressed with the method of Lagrange multipliers by applying the minimax DDP technique\cite{morimoto2003minimax} to the resulting Lagrangian function.

Starting from now we make no distinction between decision variable $x_0$ and initial state $\bar x_0$. This causes no ambiguity since we always operate with dynamically feasible state trajectories. To illustrate the underlying idea behind the algorithm we apply Bellman's principle of optimality to transform \eqref{CFTOCP} into
\begin{equation*}
    \min_{\substack{u_0\ s.t.\\ c(x_0,u_0)\leq 0}}\Big[ q(x_0,u_0) + \min_{\substack{u_1\ s.t.\\ c(f(x_0,u_0),u_1)\leq 0}}\big[q(f(x_0, u_0), u_1) + \ldots \big]\Big]
\end{equation*}
Invoking the optimality principle, i.e., using $J_0^\star(x):= p(x)$ and
\begin{equation*}
    J^{\star}_k(x)=\min_{\substack{u\ s.t.\\ c(x,u)\leq 0}}\Big[ q(x,u) + J^{\star}_{k-1}(f(x, u))\Big],
\end{equation*}
where $x \in \mathbb{R}^n$, $u \in \mathbb{R}^m$ and $k\in \{1,\ldots, N\}$, the optimisation problem \eqref{CFTOCP} is written as 
\begin{equation}\label{nested}
    J^{\star}_N(x_0)=\min_{\substack{u_0\ s.t.\\ c(x_0,u_0)\leq 0}}\Big[ q(x_0,u_0) + J^{\star}_{N-1}(f(x_0, u_0))\Big].
\end{equation}
This enables one to apply the dynamic programming method to solve the optimisation problem \eqref{CFTOCP} as a sequence of nested optimisation problems, which is the main idea behind DDP algorithm\cite{mayne1966second}. However, in this paper, instead of solving the nested problems \eqref{nested} sequentially as in \cite{murray1979constrained,yakowitz1986stagewise,lin1991differential,xie2017differential}, we replace the constrained minimisation problems with their min-max primal-dual counterparts as follows
 \begin{equation}\label{pdnested}
     J^{\star}_k(x)=\min_{u}\max_{s\geq 0}\Big[ \ell(x, u, s)+ J^{\star}_{k-1}(f(x, u))\Big],
 \end{equation}
 where $\ell(x,u,s):=q(x,u) +  s^Tc(x,u)$. 
\subsection{Primal-Dual Feasible-Interior-Point DDP}
Let $\mu$ be a strictly positive constant and assume that a current solution estimate for \eqref{KKT} is given by tuple $(\bm x, \bm u, \bm s)$, such that $x_{t+1}=f(x_{t}, u_{t})$, $c(x_t, u_t)<0$ and $s_t>0$ for $t\in\{0,\ldots,N-1\}$. The solution approach aims to improve the solution estimate by calculating the control inputs updates that minimise quadratic models of \eqref{pdnested} in the vicinity of the current trajectory through a \emph{Backward Pass}, and computing a new trajectory after a \emph{Forward Pass}. The backward and forward passes are carried out repeatedly until a convergence criterion is satisfied. 

\subsubsection{Backward pass} Define quadratic functions
\begin{equation*}
    V^{t}(x):= {\big(V_x^t\big)}^T(x-x_t) + \frac{1}{2}(x-x_t)^TV_{xx}^t(x-x_t),
\end{equation*}
where $x\in \mathbb{R}^n$ and $t\in\{0,\ldots,N\}$. Let $V_x^N=p_x(x_N)$ and $V_{xx}^N=p_{xx}(x_N)$ are the gradient and Hessian of function $p(x)$ at $x_N$ respectively, while the remaining coefficients $V_x^t$ and $V_{xx}^t$ are to be defined recursively later in the following. 

Define $Q^t(x,u,s): = \ell(x, u,s)+V^{t+1}\big(f(x,u)\big)$ for $t\in\{0,\ldots,N-1\}$, and consider its second-order variation $\delta Q^t(\delta x, \delta u, \delta s)$ around $(x_t, u_t, s_t)$
\begin{multline*}
    \delta Q^t(\delta x, \delta u, \delta s):={\begin{bmatrix}Q_x^t \\ Q_u^t \\ Q_s^t \end{bmatrix}}^T{\begin{bmatrix}\delta x \\ \delta u\\ \delta s \end{bmatrix}} 
    \\+ \frac{1}{2}
    {\begin{bmatrix} \delta x \\ \delta u\\ \delta s \end{bmatrix}}^{T}
    \begin{bmatrix}Q_{xx}^t&Q_{xu}^t&Q_{xs}^t\\
    Q_{ux}^t&Q_{uu}^t&Q_{us}^t\\
    Q_{sx}^t&Q_{su}^t&Q^t_{ss}
    \end{bmatrix} 
    {\begin{bmatrix}\delta x \\ \delta u\\ \delta s \end{bmatrix}},
\end{multline*}
with the derivatives being evaluated at $(x_t,u_t,s_t)$:
\begin{subequations}\label{dQ}
\begin{align}
        Q_s^t&=c(x_t,u_t),\ Q_{sx}^t=c_x,\ Q_{su}^t=c_u,\ Q^t_{ss}=0,\\
        Q_x^t&=\ell_x+f_x^T V_x^{t+1},\ Q_u^t=\ell_u+f_u^T V_x^{t+1},\\
        Q_{xx}^t&=\ell_{xx}+f_x^T V_{xx}^{t+1}f_x+V_{x}^{t+1}\cdot f_{xx},\\
        Q_{uu}^t&=\ell_{uu}+f_u^T V_{xx}^{t+1}f_u+V_{x}^{t+1}\cdot f_{uu},\\
        Q_{xu}^t&=\ell_{xu}+f_x^T V_{xx}^{t+1}f_u+V_{x}^{t+1}\cdot f_{xu},
\end{align}
\end{subequations}
where $\cdot$ denotes tensor contraction along an appropriate dimension.
The aim is to construct a quadratic model of $J^{\star}_{N-t}(x)$ in the vicinity of $x_t$ by providing suitable expressions for $V^t(x)$, and obtain local improvements to the decision variables. To this aim, we consider a solution to
\begin{equation}\label{local}
    \min_{\delta u}\Big[\max_{\delta s} \delta Q^t(\delta x,\delta u,\delta s)\quad \text{s.t.}\quad s_t +\delta s\geq0 \Big],
\end{equation}
where this problem is a local approximation to $$\min_{u}\max_{s\geq0} \Big[\ell(x, u,s)+V^{t+1}\big(f(x,u)\big)\Big].$$
Local minimisation of \eqref{local} with respect to $\delta u$ results in
\begin{equation} \label{eq:deltau}
Q_u^t+Q_{ux}^t \delta x +Q_{uu}^t\delta u +Q_{us}^t \delta s =0.
\end{equation}
Additionally, for $\delta s$ to be a stationary point of the inner maximization problem in \eqref{local}, it must satisfy $s_t+\delta s \geq 0$ and
\begin{equation}\label{eq:deltas}
    (s_t+\delta s)\odot(Q_s^t + Q_{sx}^t\delta x + Q_{su}^t \delta u)=0,
\end{equation}
where $\odot$ is the element-wise vector multiplication. The equation obtained from dropping all the second-order terms in \eqref{eq:deltas} and adding the perturbation vector $\bm \mu$ to its left-hand side, along with \eqref{eq:deltau}
yield the following parametric system
\begin{equation} \label{eqnsys}
\begin{bmatrix}
Q_{uu}^t & Q_{us}^t\\
S_t Q_{su}^t &  C_t
\end{bmatrix}\begin{bmatrix}
\delta u\\
\delta s
\end{bmatrix}= -\begin{bmatrix}
Q_{u}^t\\
r_t
\end{bmatrix}- \begin{bmatrix}
Q_{ux}^t\\
S_t Q_{sx}
\end{bmatrix}\delta x,
\end{equation}
where $r_t:=S_tc(x_t,u_t)+\bm \mu$, $C_t:=\operatorname{diag}[c(x_t,u_t)]$ and $S_t:=\operatorname{diag}[s_t]$. The parametric solution of \eqref{eqnsys} as a function of $\delta x$ is given by
\begin{equation}\label{updatelaws}
    \begin{bmatrix}
    \delta u\\
    \delta s
    \end{bmatrix}=\begin{bmatrix}
    \alpha_t\\
    \eta_t
    \end{bmatrix}+\begin{bmatrix}
    \beta_t\\
    \theta_t
    \end{bmatrix}\delta x,
\end{equation}
where the coefficients are established from
\begin{equation}
    \label{coefficients}
    \begin{bmatrix}
    \alpha_t & \beta_t\\
    \eta_t & \theta_t
    \end{bmatrix}=-\begin{bmatrix} 
    Q_{uu}^t & Q_{us}^t \\
    S_t Q_{su}^t & C_t
    \end{bmatrix}^{-1}\begin{bmatrix}
    Q_u^t & Q^t_{ux}\\
    r_t & S_t Q^t_{sx}
    \end{bmatrix}.
\end{equation}

Note that \eqref{coefficients} can be readily computed for $t=N-1$ as $V^N(x)$ is determined by function $p(x)$. Next, the expressions for the coefficients of $V^t(x)$ for $t\in\{0,\ldots,N-1\}$ are
\begin{equation*}\label{updateV}
    \begin{aligned}
    V_x^t&=\hat Q_x^t+\beta_t^T\hat Q_u^t+(\hat Q_{ux}^t+ \hat Q_{uu}^t\beta_t)^T\alpha_t=\hat Q_x^t+\hat Q_{xu}^t\alpha_t,\\
    V_{xx}^t&=\hat Q_{xx}^t+\hat Q_{xu}^t\beta_t +\beta_t^T \hat Q_{ux}^t+\beta_t^T \hat Q_{uu}^t\beta_t=\hat Q_{xx}^t+\hat Q_{xu}^t\beta_t,
    \end{aligned}
\end{equation*}
where
    \begin{subequations}\label{hatQ}
    \begin{align} 
        \hat Q_x^t&=Q_x^t - Q_{xs}^t C_t^{-1}r_t,\\ 
        \hat Q_u^t&=Q_u^t - Q_{us}^t C_t^{-1}r_t,\\
        \hat Q_{xx}^t&=Q_{xx}^t -Q_{xs}^t C_t^{-1}S_tQ_{sx}^t,\\ 
        \hat Q_{xu}^t&=Q_{xu}^t -Q_{xs}^t C_t^{-1}S_tQ_{su}^t,\\ 
        \hat Q_{uu}^t&=Q_{uu}^t -Q_{us}^t C_t^{-1}S_t Q_{su}^t. \label{hatQuu}
        \end{align}
    \end{subequations}
\begin{remark} \label{rem:alternative}
Expressions \eqref{hatQ} are established by replacing the standard DDP coefficients with the corresponding coefficients of the condensed system obtained from eliminating $\delta s$ in \eqref{eqnsys}.
\end{remark}
\subsubsection{Forward pass} Define the update functions 
\begin{equation}\label{updateguess}
\begin{aligned} 
        \phi_t(x)&:=u_t+\alpha_t+\beta_t (x - x_t),\\
        \psi_t(x)&:=s_t+\eta_t+\theta_t (x - x_t),
\end{aligned}    
\end{equation}
denote a new iterate by $(\bm x^+, \bm u^+, \bm s^+)$, where $\bm x^+=(x_0^+,\ldots, x_N^+)$, $\bm u^+=(u_0^+,\ldots, u_{N-1}^+)$, $\bm s^+=(s_0^+,\ldots, s_{N-1}^+)$ and $x_0^+=x_0$. For $t\in\{0,\ldots,N-1\}$ we compute
\begin{align}
    u_t^+&=\phi_t(x_t^+), \notag\\
    s_t^+&=\psi_t(x_t^+), \label{newiter} \\
    x_{t+1}^+&=f(x_t^+, u_t^+). \notag
\end{align}
\begin{remark} \label{linesearch}
Note that the strict constraints satisfaction for the new iterate should be ensured. In other words, the iterate should satisfy $c(x_t^+, u_t^+)<0$ and $s_t^+>0$ for $t\in\{0,\ldots,N-1\}$. A common approach is to employ a line-search procedure by including a step length $\gamma\in [0, 1]$ in the definition of functions as follows
\begin{equation*} \label{updateguess2}
    \begin{aligned}
        \phi_t(x, \gamma)&=u_t+\gamma \alpha_t+\beta_t (x - x_t),\\
        \psi_t(x, \gamma)&=s_t+\gamma \eta_t+\theta_t (x - x_t).
    \end{aligned}
    \end{equation*}
    Line-search procedure of this type is common for DDP methods, e.g., see  \cite[Equation (19)]{murray1979constrained} or \cite[Equation (7)]{tassa2014control}.
\end{remark}
 
\subsection{Properties of Feasible-Interior-Point DDP}
 Due to the ``direct shooting'' nature of the proposed algorithm the state trajectory vectors are not independent variables, but are functions of the initial state and control inputs, i.e., $x_{t}=f(x_{t-1},u_{t-1})$ for $t\in\{0,\ldots,N\}$. We, however, continue to write $\bm x$ to shorthand the writing.

  \begin{definition}\label{def:feas}
  A tuple $w=(\bm x, \bm u, \bm s)$ is (\textit{strictly}) \textit{primal-dual feasible}, if it satisfies (with strict inequalities) $c(x_t, u_t)\leq0$, $s_t\geq 0$ and $x_{t+1}=f(x_t,u_t)$ for all $t\in\{0,\ldots,N-1\}$.
  \end{definition}
  \begin{assump}\label{as:1}
  Matrices $\hat Q_{uu}^t$, defined as in \eqref{hatQuu} (or later in \eqref{hatQinfeasuu}), are positive definite for all $t\in\{0,\ldots,N-1\}$.
  \end{assump}
  \begin{lemma} \label{lem:1}
  Under Assumption~\ref{as:1}, if $w=(\bm x, \bm u, \bm s)$ is strictly primal-dual feasible then the linear operators
  \begin{equation*}
      P_t=\begin{bmatrix}
        Q_{uu}^t & Q_{us}^t\\
        S_t Q_{su}^t & C_t
        \end{bmatrix}^{-1},\ t\in\{0,\ldots,N-1\},
  \end{equation*}
  are continuous at $w$.
  \end{lemma}
  \begin{proof}
  Matrices $\hat Q_{uu}^t=Q_{uu}^t - Q_{us}^t C_t^{-1}S_tQ_{su}^t$ are positive definite by assumption and $C_t$ are negative definite by strict primal-dual feasibility of $w$, thus invertible. This is a sufficient condition for existence of $P_t$, which can be found using a block-wise matrix inverse formula. 
  
  The matrix inverse (when exists) is continuous in $w$ when its matrix components are continuous function of $w$. As the continuity of $Q_{uu}^{N-1}$, $Q_{us}^{N-1}$ and $C_{N-1}$ follows directly from $q(x,u)$, $c(x,u)$ and $p(x)$ being twice continuously differentiable, we readily conclude continuity of $P_{N-1}$. Next, since $f(x,u)$ is twice continuously differentiable, we establish continuity of $Q_{uu}^{N-2}$ and $P_{N-2}$. Following the recursion we can conclude continuity of $P_t$ for $t=N-3$ to $t=0$.
  \end{proof} 
  Now consider a vector-valued function $F(w, \mu)$ defined as
  \begin{equation*}
      F(w, \mu):=\big(Q_u^0, \ldots, Q_u^{N-1}, r_0, \ldots, r_{N-1}\big),
  \end{equation*}
  where $Q_u^t$ are defined as in \eqref{dQ} and $r_t=S_t c(x_t,u_t)+ \bm \mu$. Note that any primal-dual feasible tuple $w$, which is a solution of $F(w, \mu)=0$ for a given $\mu>0$, is strictly primal-dual feasible, as $r_t=0$ precludes zero components in $s_t$ and $c(x_t,u_t)$, and is a stationary point of the IPDDP algorithm, since all $\alpha_t$ and $\eta_t$ in \eqref{coefficients} are zero in this case. Moreover, such $w$ is a perturbed KKT point, i.e., a solution of the perturbed KKT system, as it is established in the following theorem, and thus can be moved arbitrary close to the (locally) optimal solution of \eqref{CFTOCP}.
  
  \begin{theorem} \label{thm:1}
  Let $w=(\bm x, \bm u, \bm s)$ be primal-dual feasible. If $w$ is a solution of $F(w, \mu)=0$ for a given $\mu>0$, then it is a perturbed KKT point, i.e., there exists $\bm \lambda = (\lambda_0,\ldots,\lambda_N)$ such that $(\bm x, \bm u, \bm \lambda, \bm s)$ satisfies \eqref{KKT}.
  \end{theorem}
  \begin{proof}
  Note that satisfaction of $\nabla_{\lambda_t}L(\bm x, \bm u, \bm \lambda, \bm s)=f(x_{t-1},u_{t-1})-x_t=0$, $c(x_t, u_t)\leq 0$ and $s_t\geq0$ for all $t\in\{0,\ldots,N\}$ follows directly from the primal-dual feasibility hypothesis.
  
  Now assume $w$ is a primal-dual solution of $F(w, \mu)=0$, meaning that $Q^t_u=0$ and $r_t=S_tc(x_t,u_t)+\bm \mu=0$ for all $t\in\{0, \ldots, N-1\}$, and note that it directly ensures satisfaction of $\bm S \bm c(\bm x, \bm u) +\bm \mu=0$ in \eqref{KKT}. Also note that 
  $\nabla_{x_N}L(\bm x, \bm u, \bm \lambda, \bm s)=0$ is satisfied when $\lambda_N=V_x^N$. For $t=N-1$ we have
  \begin{align*}
      \nabla_{x_{N-1}}L&=\ell_x+f_x^T \lambda_N -\lambda_{N-1}=Q^{N-1}_x-\lambda_{N-1}, \\
      \nabla_{u_{N-1}}L&=\ell_u+f_u^T \lambda_N=Q^{N-1}_u=0, 
  \end{align*}
  where the derivatives are evaluated at $(x_{N-1}, u_{N-1}, s_{N-1})$. Proceeding with $\lambda_t=V^{t}_x=Q^{t}_x$ we obtain $(\bm x, \bm u, \bm \lambda, \bm s)$, where $\bm \lambda =(\lambda_0,\ldots,\lambda_N)$, which satisfies system \eqref{KKT}. 
  \end{proof}  
  Moreover, once $w$ is sufficiently close to a solution of $F(w, \mu)=0$, the IPDDP iterates converge to the solution at a quadratic rate. This is demonstrated in the following theorem.
  \begin{theorem} \label{thm:2}
   Let $w=(\bm x, \bm u, \bm s)$ and $w^+=(\bm x^+, \bm u^+, \bm s^+)$, defined by \eqref{newiter}, are strictly primal-dual feasible, and $\mu>0$. There exist $M\geq 0$ and $\varepsilon>0$ such that if $\|w-w^\star\|\leq\varepsilon$, then
  \begin{equation}\label{thm1:eq}
    \begin{aligned}
      \| w^+ - w^\star\| &\leq M \| w - w^\star\|^2,\\
      \| w^+ - w^\star\| &< \| w - w^\star\|,
    \end{aligned}
  \end{equation}
  where $w^\star=(\bm x^\star, \bm u^\star, \bm s^\star)$ is a primal-dual feasible solution of $F(w, \mu)=0$.
  \end{theorem}
  \begin{proof}
  Consider functions defined by $Q_u^t$ and $r_t$, and note that they are differentiable, as all the terms involved are differentiable. Then, by the Taylor theorem there exist functions $h_t(w, w^\star)$ and $g_t(w, w^\star)$ such that
    \begin{align}\label{taylor}
          Q_u^t\big\rvert_{w^\star}= Q_u^t + Q_{ux}^t (x_t^\star-x_t) + Q_{uu}^t (u_t^\star - u_t) \\+ Q_{us}^t (s_t^\star - s_t)+ h_t(w,w^\star)&=0,\notag\\
          r_t\big\rvert_{w^\star}= r_t+S_t Q_{sx}^t (x_t^\star - x_t) + S_t Q_{uu}^t (u_t^\star - u_t)  \notag\\+ C_t (s_t^\star - s_t)+ g_t(w,w^\star)&=0,\notag
    \end{align}
  where we used that $w^\star$ is a solution of $F(w,\mu)=0$, and the norms of the residual functions are bounded:
  \begin{align*}
      \|h_t(w,w^\star)\|&\leq H_t \|w-w^\star\|^2,\\
      \|g_t(w,w^\star)\|&\leq G_t \|w-w^\star\|^2,
  \end{align*}
  where $H_t$ and $G_t$ for $t\in\{0,\ldots,N-1\}$ are some constants.
  Denote $\Delta x_t = x_t^+ - x_t$, $\Delta u_t = u_t^+ - u_t$, $\Delta s_t = s_t^+ - s_t$ for $t\in\{0,\ldots,N-1\}$, and notice that $(\Delta x_t,\Delta u_t,\Delta u_t)$ is a solution of system \eqref{eqnsys}, as it belongs to parametric family \eqref{updatelaws}:
  \begin{align*}
      \Delta u_t&= u_t^+ - u_t = \phi (x_t^+) - u_t= \alpha_t + \beta_t \Delta x_t,\\
      \Delta s_t&= s_t^+ - s_t = \psi (x_t^+) - s_t= \eta_t + \theta_t \Delta x_t.
  \end{align*}
 Now, by adding and subtracting $x_t^+$, $u_t^+$, $s_t^+$ in the appropriate parentheses of \eqref{taylor}, and using the fact that $(\Delta x_t,\Delta u_t,\Delta u_t)$ is a solution of \eqref{eqnsys}, we have
  \begin{align*}\label{taylorsys}
      Q_{ux}^t (x_t^\star-x_t^+) + Q_{uu}^t (u_t^\star -u_t^+)+ Q_{us}^t (s_t^\star -s_t^+)\\+ h_t(w,w^\star)&=0,\\
      S_t Q_{sx}^t (x_t^\star -x_t^+) + S_t Q_{su}^t (u_t^\star -u_t^+) + C_t (s_t^\star -s_t^+)\\+ g_t(w,w^\star)&=0,
  \end{align*}
  which can be rewritten using operator $P_t$ as
  \begin{equation*}
    \begin{bmatrix}
    u_t^\star -u_t^+\\
    s_t^\star -s_t^+
    \end{bmatrix}=-P_t\begin{bmatrix}
    h_t(w,w^\star)\\
    g_t(w,w^\star)
    \end{bmatrix}-P_t \begin{bmatrix}
    Q^t_{ux}\\
    S_t Q^t_{sx}
    \end{bmatrix} (x_t^\star-x_t^+).
  \end{equation*}
  Since $P_0$ is a continuous function of $w$, and $x_0^\star=x_0^+=x_0$, for $t=0$ this gives 
  \begin{align*}
      \big\|[u_0^\star - u_0^+, s_0^\star - s_0^+]\big\| & \leq \big\|P_0\big\|\big\|(h_0(w,w^\star),g_0(w,w^\star))\big\|\\&\leq M_0\big\|w-w^\star\big\|^2,
  \end{align*}
  where $M_0=(H_0+G_0)\max_{w\in\Omega}\|P_0\|$, and the operator norm $\|P_0\|$ (induced by Euclidian vector norm) of continuous linear operator $P_0$ is bounded on a compact subset $\Omega$ of all strictly primal-dual feasible $w$, which can be chosen to be sufficiently big, i.e., such that it contains set of all strictly primal-dual feasible $w: \| w - w^\star\|\leq\varepsilon$ for any $\varepsilon \leq \bar \varepsilon$, where $\bar \varepsilon>0$ is a constant.
  By using Taylor formula for $f(x_0^\star,u_0^\star)$, and bounding the residuals in $\|w-w^\star\|\leq \varepsilon$ we get:
  \begin{align*}
      \|x_1^\star - x_1^+\| &= \big\|f(x_0^\star,u_0^\star) - f(x_0^+,u_0^+)\big\| \\
      &\leq \big\|f_x^T (x_0^\star-x_0^+) + f_u^T (u_0^\star-u_0^+)\big\| \\ &\quad + K_1 \big\| (x_0^\star-x_0^+, u_0^\star-u_0^+)\big\|^2\\
      &\leq K_2 \big\| (x_0^\star-x_0^+, u_0^\star-u_0^+)\big\|\leq K_3 \| w-w^\star\| ^2,
  \end{align*}
  for some constants $K_1,K_2,K_3\geq0$. Continuing by induction, there are some constants $M_t\geq0$ for $t\in\{0,\ldots,N-1\}$, such that for a positive $\varepsilon\leq \bar \varepsilon$ and all strictly primal-dual feasible $w:\|w-w^\star\|\leq \varepsilon $ we have:
  $$\big\|(x_t^\star - x_t^+, u_t^\star - u_t^+, s_t^\star - s_t^+)\big\| \leq M_t\big\|w-w^\star\big\|^2,$$
  which proves the first inequality in \eqref{thm1:eq} by choosing $M=M_1+M_2+\ldots+M_N$:
  \begin{align*}
      \big\|w^\star - w^+\big\| & \leq\sum_{t=1}^N \big\|(x_t^\star - x_t^+, u_t^\star - u_t^+, s_t^\star - s_t^+)\big\|\\& \leq M \big\|w-w^\star\big\|^2.
  \end{align*} 
  The second inequality in \eqref{thm1:eq} is obtained from the first one by pickings $\varepsilon$ sufficiently small, i.e., such that $\varepsilon<1/ M$ in addition to $\varepsilon\leq\bar \varepsilon$. Now, for all strictly primal-dual feasible $w$ such that $\|w-w^\star\|\leq \varepsilon$ we additionally have $
      \big\|w^+-w^\star\big\| \leq \varepsilon M  \big\|w-w^\star\big\|<\big\|w-w^\star\big\|$.
  \end{proof}

  \subsection{Primal-Dual Infeasible-Interior-Point DDP}
  In the previous sections, we assumed that one has access to a strictly primal-dual feasible initial solution guess, which might be quite restrictive in practice. Here we propose an extension where this assumption is relaxed. In this section we explain what should be modified in the Feasible-IPDDP algorithm, and how the theoretical results translate to a new algorithm. 
    \subsubsection{Backward pass} 
    We use the same definitions for $V^{t}(x)$ and $Q^t(x)$, and follow the same steps to obtain the first equation of system \eqref{eqnsys}. Next, however, we introduce slack variables $y_t\geq 0$ for $t\in\{0,\ldots,N-1\}$ to transform the inequality constraints into equalities, i.e., $c(x_t, u_t) + y_t = 0$, and use the perturbed complementarity condition, formulated in terms of the dual and slack variables, i.e., $S_ty_t-\bm \mu=0$. 
    Finally, we proceed by setting the first order Taylor expansion for $c(x_t, u_t) + y_t$ and $S_ty_t-\bm \mu$ to zero. In this way, the parametric system of equations becomes
    \begin{equation}\label{infeas:sys}
    \begin{bmatrix} Q_{uu}^t & Q_{us}^t & 0\\
                  Q_{su}^t & 0    & I\\
                  0      & Y_t  & S_t
    \end{bmatrix}
    \begin{bmatrix} \delta u\\
                  \delta s\\
                  \delta y
    \end{bmatrix}
    =-\begin{bmatrix} Q_u^t\\
                  r^p_t\\
                  r^d_t
    \end{bmatrix}-\begin{bmatrix} Q_{ux}^t\\
                  Q_{sx}^t\\
                  0
    \end{bmatrix}\delta x,
    \end{equation}
    where $r_t^p:=c(x_t,u_t)+y_t$, $r_t^d:=S_ty_t - \bm \mu$ and $Y_t:=\operatorname{diag}[y_t]$. 
    Note that elimination of $\delta y = -r^p_t - Q_{su}\delta u - Q_{sx}\delta x$ from the system leads to a reduced system similar to \eqref{eqnsys}
    \begin{equation*}
    \begin{bmatrix} Q_{uu}^t & Q_{us}^t\\
                  S_t Q_{su}^t & -Y_t\\
    \end{bmatrix}
    \begin{bmatrix} \delta u\\
                  \delta s
    \end{bmatrix}
    =-\begin{bmatrix} Q_u^t\\
                  \hat r_t
    \end{bmatrix}-\begin{bmatrix} Q_{ux}^t\\
                  S_t Q_{sx}^t
    \end{bmatrix}\delta x,
    \end{equation*}
    where $\hat r_t:=S_t r^p_t-r^d_t$.
    Solving the parametric system
    \begin{equation*}
    \label{infeas:coefficients}
    \begin{bmatrix}
    \alpha_t & \beta_t\\
    \eta_t & \theta_t\\
    \chi_t & \zeta_t
    \end{bmatrix}=-\begin{bmatrix} Q_{uu}^t & Q_{us}^t & 0\\
                  Q_{su}^t & 0    & I\\
                  0      & Y_t  & S_t
    \end{bmatrix}^{-1}\begin{bmatrix}
    Q_u^t & Q^t_{ux}\\
    r_t^p & Q^t_{sx}\\
    r_t^d & 0
    \end{bmatrix},
    \end{equation*}
    yields the new coefficients
    \begin{subequations}\label{hatQinfeas}
    \begin{align}
        \hat Q_x^t &=Q_x^t + Q_{xs}^t Y_t^{-1}\hat r_t,\\ 
        \hat Q_u^t &=Q_u^t + Q_{us}^t Y_t^{-1}\hat r_t,\\
        \hat Q_{xu}^t&=Q_{xu}^t +Q_{xs}^t Y_t^{-1}S_tQ_{su}^t,\\ 
        \hat Q_{xx}^t&=Q_{xx}^t +Q_{xs}^t Y_t^{-1}S_tQ_{sx}^t,\\ 
        \hat Q_{uu}^t&=Q_{uu}^t +Q_{us}^t Y_t^{-1}S_t Q_{su}^t. \label{hatQinfeasuu}
    \end{align}
    \end{subequations}
    \subsubsection{Forward pass}
    Define the update functions 
    \begin{align*}
        \phi_t(x)&=u_t + \alpha_t + \beta_t (x - x_t),\\
        \psi_t(x)&=s_t+\eta_t + \theta_t (x -x_t),\\
        \xi_t(x)&=y_t+\chi_t + \zeta_t(x -x_t),
    \end{align*}
    and denote a new solution guess by $(\bm x^+, \bm u^+, \bm s^+, \bm y^+)$, where $\bm x^+=(x_0^+,\ldots, x_N^+)$, $\bm u^+=(u_0^+,\ldots, u_{N-1}^+)$, $\bm s^+=(s_0^+,\ldots, s_{N-1}^+)$, $\bm y^+=(y_0^+,\ldots, y_{N-1}^+)$ and $x_0^+=x_0$. For $t\in\{0,\ldots,N-1\}$ we compute
    \begin{equation}\label{infeas:iter}
      \begin{aligned} 
        u_t^+&=\phi_t(x_t^+),\\
        s_t^+&=\psi_t(x_t^+),\\
        y_t^+&=\xi_t(x_t^+),\\
        x_{t+1}^+&=f(x_t^+, u_t^+). 
    \end{aligned}  
    \end{equation}
    \begin{remark}
    Note that now only positivity of $s_t^+$ and $y_t^+$ must be ensured. Satisfaction of $c(x,u)\leq 0$ is only guaranteed in the limit of the convergence, which is a common property for the so-called \textit{``infeasible''} methods. This potentially facilitates algorithm implementation, but an appropriate line-search procedure is still required in practical implementations, see \cite{nocedal2006numerical} for details.\end{remark}
  \subsection{Properties of Infeasible-Interior-Point DDP} 
  Now we add slack variables $\bm y=(y_0,\ldots,y_{N-1})$ into the tuple $w = (\bm x, \bm u, \bm s, \bm y)$, and provide the definition of \textit{dual feasibility} as follows.
  \begin{definition}\label{def:infeas}
  A tuple $w=(\bm x, \bm u, \bm s, \bm y)$ is (\textit{strictly}) \textit{dual feasible}, if it satisfies (with strict inequalities) $s_t\geq 0$, $y_t\geq 0$ and $x_{t+1}=f(x_t,u_t)$ for all $t\in\{0,\ldots,N-1\}$.
  \end{definition}
  Consider a vector-valued function $F(w, \mu)$ defined as
  \begin{equation*}
      F(w, \mu):=\big(Q_u^0, \ldots, Q_u^{N-1}, r_0^p, \ldots, r_{N-1}^p, r_0^d, \ldots, r_{N-1}^d\big),
  \end{equation*}
  where $r_t^p:=c(x_t,u_t) + y_t$ and $r_t^d:= S_ty_t -\bm \mu $. 
  \begin{lemma} \label{lem:2}
  Under Assumption~\ref{as:1}, if $w = (\bm x,\bm u,\bm s,\bm y)$ is strictly dual feasible then the linear operators
  \begin{equation*}
    P_t=\begin{bmatrix} Q_{uu}^t & Q_{us}^t & 0\\
                      Q_{su}^t & 0    & I\\
                      0      & Y_t  & S_t
        \end{bmatrix}^{-1},\ t\in\{0,\ldots,N-1\},
  \end{equation*}
  are continuous at $w$.
  \end{lemma}
  \begin{proof}
  The proof is similar to that of Lemma~\ref{lem:1}.
  \end{proof} 
  \begin{corollary}
  Let $w=(\bm x, \bm u, \bm s, \bm y)$ be dual feasible. If $w$ is a solution of $F(w, \mu)=0$ for a given $\mu>0$, then it is a perturbed KKT point, i.e., there exists $\bm \lambda = (\lambda_0,\ldots,\lambda_N)$ such that $(\bm x, \bm u, \bm \lambda, \bm s)$ satisfies \eqref{KKT}.
  \end{corollary}
  \begin{proof}
  Note that $r_t^d=c(x_t, u_t) +y_t=0$ for $t\in\{0,\ldots,N\}$ ensures primal-dual feasibility. With this in mind one can follow the same steps as in the proof of Theorem~\ref{thm:1}. 
  \end{proof}
  \begin{theorem}
  Let $w=(\bm x, \bm u, \bm s, \bm y)$ and $w^+=(\bm x^+, \bm u^+, \bm s^+, \bm y^+)$, defined by \eqref{infeas:iter}, are strictly dual feasible, and $\mu>0$. There exist $M\geq 0$ and $\varepsilon>0$ such that if $\|w-w^\star\|\leq\varepsilon$, then
  \begin{subequations}\label{lem1:eq}
    \begin{align}
      \| w^+ - w^\star\| &\leq M \| w - w^\star\|^2, \\
      \| w^+ - w^\star\| &< \| w - w^\star\|,
    \end{align}
  \end{subequations}
  where $w^\star=(\bm x^\star, \bm u^\star, \bm s^\star, \bm y^\star)$ is a dual feasible solution of $F(w, \mu)=0$.
  \end{theorem}
  \begin{proof}
  The proof follows the same steps as the proof of Theorem~\ref{thm:1}. We consider differentiable functions defined by $Q_u^t$, $r_t^p$ and $r_t^d$, and use its first-order Taylor expansion at $w^\star$:
  \begin{align*}\label{inf:taylor}
      Q_u^t\big\rvert_{w^\star}= Q_u^t + Q_{ux}^t (x_t^\star-x_t) + Q_{uu}^t (u_t^\star - u_t) \\+ Q_{us}^t (s_t^\star - s_t)+ h_t(w,w^\star)&=0,\\
      r_t^p\big\rvert_{w^\star}= r_t^p + Q^t_{sx} (x_t^\star-x_t) + Q^t_{su}(u_t^\star-u_t) \\+ (y_t^\star-y_t) + g_t(w,w^\star)&=0,\\
      r_t^d\big\rvert_{w^\star}= r_t^d+ Y_t (s_t^\star - s_t)+S_t (y_t^\star - y_t) \\+ k_t(w,w^\star)&=0,
  \end{align*}
  and the norms of the residuals are bounded
  \begin{align*}
  \|h_t(w,w^\star)\|&\leq H_t \|w-w^\star\|^2,\\ \|g_t(w,w^\star)\| &\leq G_t \|w-w^\star\|^2,\\ 
  \|k_t(w,w^\star)\| & \leq K_t \|w-w^\star\|^2,
  \end{align*}
  where $H_t$, $G_t$ and $K_t$ for $t\in\{0,\ldots,N-1\}$ are some constants.
  As before, we use $\Delta x_t = x_t^+ - x_t$, $\Delta u_t = u_t^+ - u_t$, $\Delta s_t = s_t^+ - s_t$, $\Delta y_t = y_t^+ - y_t$ for $t\in\{0,\ldots,N\}$, and note that $(\Delta x_t,\Delta u_t,\Delta u_t, \Delta y_t)$ is a solution of system \eqref{infeas:sys}. Thus,
  \begin{align*}
      Q_{ux}^t (x_t^\star-x_t^+) + Q_{uu}^t (u_t^\star - u_t^+) + Q_{us}^t (s_t^\star - s_t^+)\\+ h_t(w,w^\star)&=0,\\
      Q^t_{sx} (x_t^\star-x_t^+) + Q^t_{su}(u_t^\star-u_t^+) + (y_t^\star-y_t^+) \\+ g_t(w,w^\star)&=0,\\
      Y_t (s_t^\star - s_t^+)+S_t (y_t^\star - y_t^+)  + k_t(w,w^\star)&=0,
  \end{align*}
  and using the new definition of operators $P_t$:
  \begin{equation*}
    \begin{bmatrix}
    u_t^\star -u_t^+\\
    s_t^\star -s_t^+\\
    y_t^\star -y_t^+
    \end{bmatrix}=- P_t\begin{bmatrix}
    h_t(w,w^\star)\\
    g_t(w,w^\star)\\
    k_t(w,w^\star)
    \end{bmatrix}- P_t \begin{bmatrix}
    Q^t_{ux}\\
    Q^t_{sx}\\
    0
    \end{bmatrix} (x_t^\star-x_t^+).
  \end{equation*}
  From here the remaining proof is straightforward: we start with $t=0$ and proceed iteratively to $t=N-1$ as it is described in the proof of Theorem~\ref{thm:1}. 
  \end{proof}
  \section{Numerical simulations}\label{sec:examples}
  Here we compare implementations of the proposed Interior-Point DDP (IPDDP) algorithms\footnote{ https://github.com/xapavlov/ipddp} with implementations of Control-Limited DDP (CLDDP)\cite{YuvalMatlab} and log-barrier DDP. 
  
  The wall-clock time required for solving problems using these methods depends mainly on the number of iterations and time spent per iteration. While we focus on the total number of iterations required for algorithms to converge to a (locally) optimal solution, we briefly note that the per-iteration complexity for each algorithm, due to their nature, is different. Specifically, CLDDP require solving $N$ box-QPs (of size $m$ with $l$ constraints) per iteration, IPDDP and log-barrier DDP require solving $N$ linear systems of equations (of ``effective'' size $m$) per iteration. Thus, one expects that each iteration of the CLDDP takes longer than that of its competitors. However, a fair wall-clock time comparison between methods is beyond the scope of this paper.
  
  To ensure convergence of the proposed algorithms to the perturbed KKT points we rely on the line-search (see Remark~\ref{linesearch}), step filter\cite{fletcher2002nonlinear} and regularisation of type $Q_{uu}^t+\gamma I$. The regularisation is essential when Assumption~\ref{as:1} is violated, what commonly occurs at the early stages of the optimisation routine, but holds later on (in the considered examples). Perturbation $\mu$ is initialised as $\mu \leftarrow J(\bm x, \bm u)/(N l)$, where $J(\bm x, \bm u)$ is the value of the objective function. Perturbation is reduced as $\mu \leftarrow \min(\mu/\kappa, \mu^{1.2})$, where we pick the \emph{reduction factor} $\kappa=5$ (unless stated otherwise),  every time $\| F(w,\mu)\|_{\infty}$ is sufficiently small\footnote{Since the infinity norms of the objective function gradient and constraints violations at a step are proportional with $\mu$, it can be used as a surrogate for the progress of the algorithm.}, i.e., less than than $0.2\mu$ in our implementation.
  In all trials we use random initial control inputs, which are sampled from interval $[-0.01, 0.01]$ at uniform, and conduct 40 trials for each experiment. We use the following definition for the logarithmic optimality error
  $$E_J:= \log_{10}\big[J(\bm x, \bm u) - J(\bm x^{\star}, \bm u^{\star})\big],$$ 
  where $J(\cdot, \cdot)$ is the objective function evaluated at a current solution guess $(\bm x, \bm u)$ or at a (numerically obtained) locally optimal solution $(\bm x^{\star}, \bm u^{\star})$.

  \subsection{Inverted pendulum}
\begin{figure}[t]\vspace{-7.5pt}
  \begin{center}
  \input{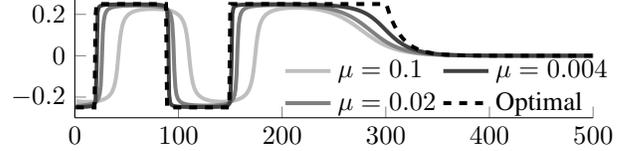}
  \end{center}
  \caption{Inverted pendulum: control inputs $u_t$ on y-axis vs time-index $t$ on x-axis.}
  \label{fig:2}
  \end{figure}

  Consider a task of stabilising the inverted pendulum
  \begin{equation*}
      f(x,u) =\begin{bmatrix}
      \varphi+h \omega\\ 
      \omega+h\sin(\varphi)+h u 
      \end{bmatrix},
  \end{equation*}
  where state $x=[\varphi\ \omega]^T$ ($\varphi$ is angle and $\omega$ is angular velocity), $u$ is the control input, $h=0.05$; initial state is $x_0=[-\pi\ 0]^T$, control constraints are $-0.25\leq u\leq 0.25$, and $N=500$. We choose the quadratic stage and terminal costs as
  \begin{equation*}
   q(x,u) = 0.025(\varphi^2 + \omega^2 + u^2),\ p(x)= 5(\varphi^2 + \omega^2).
  \end{equation*}
  \begin{figure}[t]
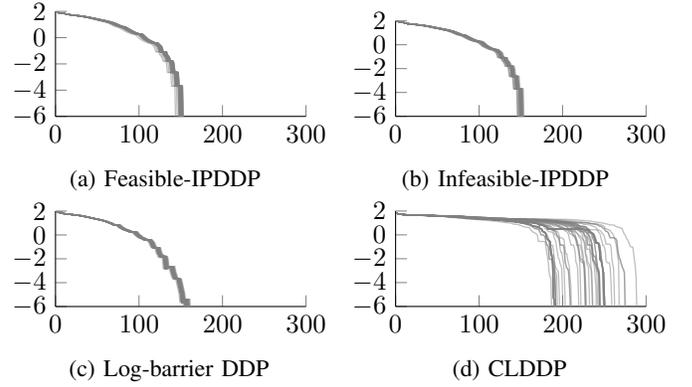
\vspace{-13.5pt}
    \subfloat[ ][Feasible-IPDDP ]{\centering\input{invpend_feas.tex}
    \label{fig1:subfig1}}
    \subfloat[ ][Infeasible-IPDDP ]{\centering\input{invpend_infeas.tex}
    \label{fig1:subfig2}}
    
    \subfloat[ ][Log-barrier DDP]{\centering\input{invpend_logbar.tex}
    \label{fig1:subfig3}}
    \subfloat[ ][CLDDP]{\centering\input{invpend_clddp.tex}
    \label{fig1:subfig4}}
    \caption{Inverted pendulum: logarithmic optimality error $E_J$ on y-axis vs iteration number on x-axis.}\label{fig:1}\vspace{-10pt}
\end{figure}
  Fig.~\ref{fig:1} shows that the log-barrier DDP, Feasible- and Infeasible-IPDDP consistently convergence to the optimal solution in about 150 iterations. On the other hand, most runs of Control-Limited DDP (CLDDP) take more than 200 iterations to converge. We believe that the bang-bang nature of the optimal control complicates the correct identification of active and inactive constraints for CLDPP algorithm. This potential issue is mitigated for the other algorithms by sufficiently perturbing the problem, and following the central path, i.e., a trace of stationary points defined for different values of perturbations, see Fig.~\ref{fig:2}.
  \begin{figure}[t]\vspace{-10pt}
  \begin{center}
  \input{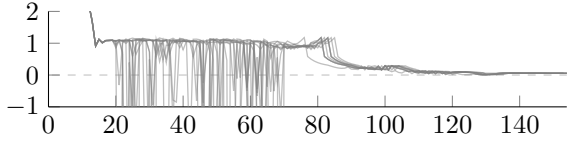}
  \end{center}
  \caption{Inverted pendulum: the smallest eigenvalue of $\hat Q_{uu}^t$ on y-axis vs iteration number on x-axis (10 trials of Feasible-IPDDP).}\vspace{-10pt}
  \label{fig:3}
  \end{figure} 
  \begin{figure}[t]\vspace{-7.5pt}
  \begin{center}
  \input{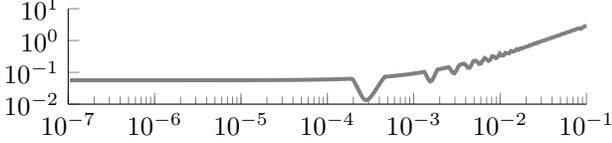}
  \end{center}
  \caption{Inverted pendulum: the smallest eigenvalue of $\hat Q_{uu}^t$ on y-axis vs perturbation $\mu$ on x-axis.}\vspace{-10pt}
  \label{fig:4}
  \end{figure} 
  
  Fig.~\ref{fig:3} demonstrates that Assumption~\ref{as:1} is commonly violated at the early stage of the optimisation routine, i.e., $\hat Q_{uu}^t$ is not positive definite, and the regularisation has to applied. Later in this example, after about 70 iterations, Assumption~\ref{as:1} is always satisfied, and the quadratic convergence rate steps in. This is because $\hat Q_{uu}^t\succ0$ in a neighbourhood (by continuity) of the central path: see Fig.~\ref{fig:4}, which depicts the smallest eigenvalues (over all $t\in\{0,\ldots,N\}$) of $\hat Q_{uu}^t$ evaluated at the perturbed KKT points along the central path.

  \subsection{Car parking problem}
  Next we consider a car parking problem as in \cite{tassa2014control}. The dynamics of a car with state $x=[r_x\ r_y\ \varphi\ v]^T$ ($r_x$ and $r_y$ are the x- and y-coordinates, $\varphi$ is the car's heading and $v$ is its velocity) and input $u=[w\ a]^T$ ($w$ and $a$ are the front wheels' steering angle and acceleration respectively) is
  \begin{equation*}
    f(x,u)=
    \begin{bmatrix}r_x+b(v,w) \cos(\varphi)\\
                   r_y+b(v,w) \sin(\varphi)\\
            \varphi +\sin^{-1}\big( \frac{hv}{d}\sin(w)\big)\\
            v + ha
      \end{bmatrix}, 
  \end{equation*}
  where $b(v,w) =d+h v\cos(\omega) -\sqrt{d^2-h^2 v^2\sin^2(w)}$, $h=0.03$ and $d=2$ is a distance between the front and back axles of a car; initial state is $x_0=[1\ 1\ 3\ \pi/2\ 0]^T$, control constraints are $-0.5\leq w\leq 0.5$ and $-2\leq a\leq 2$, and $N=500$. The cost functions are
  \begin{align*}
      q(x,u)= 0.01\big(H(r_x,0.1)+H(r_y,0.1) + w^2 + 0.01a^2\big),\\
      p(x) = H(r_x,0.1)+H(r_y,0.1)+H(\varphi,0.01)+H(v,0.1),
  \end{align*}
  where $H(y, z) =\sqrt{y^2 + z^2} - z$.
    \begin{figure}[t]
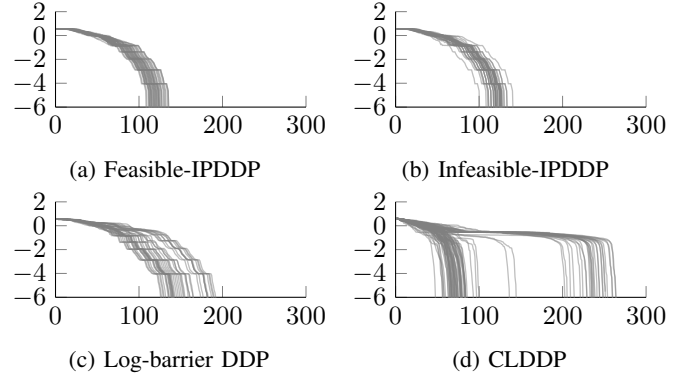
\vspace{-10pt}
    \subfloat[ ][Feasible-IPDDP ]{\centering\input{car_feas.tex}
    \label{fig5:subfig1}}
    \subfloat[ ][Infeasible-IPDDP ]{\centering\input{car_infeas.tex}
    \label{fig5:subfig2}}
    
    \subfloat[ ][Log-barrier DDP]{\centering\input{car_logbar.tex}
    \label{fig5:subfig3}}
    \subfloat[ ][CLDDP]{\centering\input{car_clddp.tex}
    \label{fig5:subfig4}}
    \caption{Car-parking problem: logarithmic optimality error $E_J$ on y-axis vs iteration number on x-axis.}\label{fig:5}\vspace{-10pt}
    \end{figure}

\begin{figure}[t]
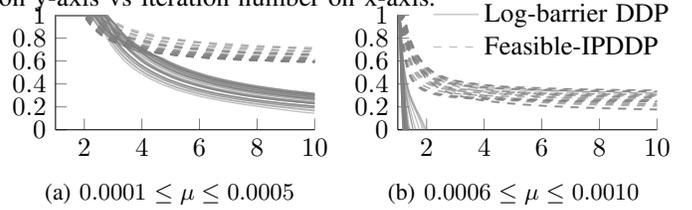
\vspace{-13.5pt}
    \subfloat[ ][$0.0001\leq\mu\leq0.0005$]{\centering\input{stepvsmu2.tex}
    \label{fig6:subfig1}}
    \subfloat[ ][$0.0006\leq\mu\leq0.0010$]{\centering\input{stepvsmu.tex}
    \label{fig6:subfig2}}
    \caption{Car-parking problem: the maximum accepted stepsize on y-axis vs the perturbation reduction factor $\kappa$ on x-axis for two different intervals of the central path.}\label{fig:6}
   \end{figure}

    In many cases CLDDP converges faster than its competitors (see Fig.~\ref{fig:5}), but presence of very distinct locally optimal solutions greatly influences its performance. IPDDP and log-barrier algorithms follow the central path, thus behave in a more predictable manner. The main advantage of the proposed algorithms compared to the log-barrier variant is their improved ability to follow the central path. To demonstrate this, we solve the perturbed KKT system for perturbation $\mu$ sampled from a certain interval, and investigate the maximum stepsizes that the  Feasible-IPDDP and log-barrier DDP methods can take immediately after the reduction of the perturbation paramter. As one can see from Fig.~\ref{fig:6}, the log-barrier DDP is more sensitive to the perturbation reduction when $\mu\in[0.0001, 0.0005]$, and takes vanishingly small steps in the case where $\mu\in[0.0006, 0.0010]$. Although such severe incidents might not commonly take place, it is known that the Newton's method applied to the log-barrier problems, in contrast with the primal-dual approach, generally suffers from the step-scaling issues occurring after the reduction of the perturbation\cite{wright1995pure}. 

    \subsection{Unicycle motion control}
\begin{figure}[t]\vspace{-10pt}
  \begin{center}
  \input{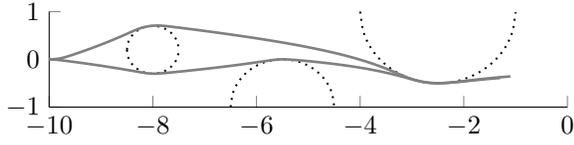}
  \end{center}
  \caption{Unicycle motion control: two distinct locally optimal trajectories (solid lines) with $r_y$ on y-axis and $r_x$ on x-axis; the boundary of obstacles depicted as dotted circles.}\label{fig:7}\vspace{-10pt}
  \end{figure}

    Here we consider a unicycle model
    $$f(x,u)= \begin{bmatrix}r_x + h v \cos(\varphi)\\r_y + hv \sin(\varphi) \\ \varphi+h u \end{bmatrix},$$ where $x=[r_x\ r_y\ \varphi]^T$ is the state vector ($r_x$ and $r_y$ are the x- and y-coordinates, $\varphi$ is the heading) and $u$ is the control input, $h=0.1$ and $v=1.5$; the initial state is $x_0=[-10\ 0\ 0]^T$ and $N=600$. We use the quadratic stage and terminal costs
    \begin{equation*}
        q(x,u)=0.1(r_x^2 + r_y^2 +\varphi^2 + 0.1 u^2),\ p(x)=0.1(r_x^2 + r_y^2 +\varphi^2). 
    \end{equation*}
    The input and state constraints are chosen as follows
    \begin{align*}
        -1.5\leq u \leq 1.5,\
        -1\leq r_y \leq 1,\ \|r_x+5.5, r_y+1\|\geq 1,\\
         \|[r_x+8,\ r_y-0.2]\|\geq 0.5,\
         \|[r_x+2.5,\ r_y-1]\|\geq 1.5.
    \end{align*}
     Fig.~\ref{fig:7} depicts two distinct locally optimal trajectories; guessing a feasible solution for this problem is not trivial. Next we will compare the relaxed logarithmic barrier strategy\cite{hauser2006barrier} implemented as DDP algorithm and Infeasible-IPDDP.
    \begin{figure}[t]
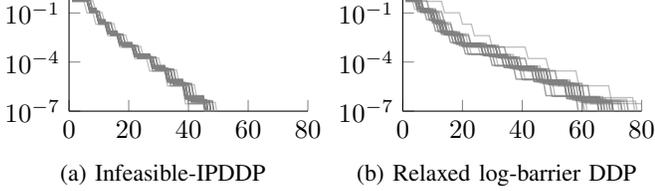
\vspace{-7.5pt}
    \subfloat[ ][Infeasible-IPDDP]{\centering\input{plane_ipddp.tex}\label{fig8:subfig1}}
    \subfloat[ ][Relaxed log-barrier DDP]{\centering\input{plane_logbar.tex}\label{fig8:subfig2}}
    \caption{Unicycle motion control: perturbation $\mu$ on y-axis vs iteration number on x-axis.}\label{fig:8}\vspace{-10pt}
    \end{figure}
   
   For the relaxed log-barrier DDP method we use the following penalty on the inequality constraints components
   \begin{equation*}
   \beta_\delta(z) =\begin{cases} 
      -\log z  & z>\delta,\\
      \frac{1}{2}\Big[\Big( \frac{z-2\delta}{\delta}\Big)^2-1\Big] - \log \delta & z\leq \delta,
   \end{cases}
   \end{equation*}
   and update the relaxation parameter as $\delta\leftarrow\mu$. For this experiment we sample initial perturbation $\mu$ from interval $[0.5, 1]$ at uniform. Fig.~\ref{fig:8} illustrates the convergence of the methods, where IPDDP outperforms its competitor. Note that $\mu$ is decreased only when the corresponding perturbed (or barrier) problem is solved with sufficient accuracy, i.e., when the norm of $F(\bm w, \mu)$ for Infeasible-IPDDP or the gradient of the penalised objective function for the relaxed log-barrier DDP are less than $0.2 \mu$, thus acting as an indicator of progress.
    
    \section{Conclusions}\label{sec:con}
   Two variants of Primal-Dual Interior-Point DDP algorithms, namely Feasible- and Infeasible-IPDDP have been proposed in this paper. The former operates with strictly feasible iterations, while the latter allows to start with infeasible solution guess, and ensures constraint satisfaction upon convergence. We prove the local quadratic convergence of the algorithms to its stationary points, and show that the stationary points of both algorithms satisfy the (perturbed) system of first-order conditions for optimality. The numerical simulations on three examples demonstrate the ability of the proposed algorithms to converge to the locally optimal solutions of the discrete-time optimal control problems with nonlinear dynamics and constraints, often being superior to its closest competitors in the total number (and variability) of iterations required to converge. 
    
    A future research direction is to establish the global convergence of IPDDP methods to the local optima. Also, we believe that the numerical performance can be further improved by implementing adaptive schemes for selecting the perturbation parameter and/or predictor-corrector steps. Other promising extensions include generalisation to differentiable manifolds.


\bibliographystyle{IEEEtran}
\small{
\bibliography{ms}  
}
\end{document}